\documentclass[12pt]{article}

\usepackage{graphicx}
\begin{document}

\newcommand{\siml}{\stackrel{<}{\sim}}
\newcommand{\simg}{\stackrel{>}{\sim}}
\newcommand{\lleq}{\stackrel{<}{=}}

\baselineskip=1.333\baselineskip


%
\begin{center}
{\large\bf
Dynamics of Information Entropies
in Nonextensive Systems
} 
\end{center}

\begin{center}
Hideo Hasegawa
\footnote{E-mail address:  hideohasegawa@goo.jp}
\end{center}

\begin{center}
{\it Department of Physics, Tokyo Gakugei University,  \\
Koganei, Tokyo 184-8501, Japan}
\end{center}
\begin{center}
({\today})
\end{center}
\thispagestyle{myheadings}

\begin{abstract}
We have discussed dynamical properties of the Tsallis entropy 
and the generalized Fisher information
in nonextensive systems described by the Langevin model 
subjected to additive and multiplicative noise.
Analytical expressions for the time-dependent
Tsallis entropy and generalized Fisher information
have been obtained with the use of the 
$q$-moment approach to the Fokker-Planck equation
developed in a previous study
[H. Hasegawa, Phys. Rev. E {\bf 77}, 031133 (2008)].
Model calculations of the information entropies  
in response to an applied pulse and sinusoidal inputs
have been presented. 
\end{abstract}

\vspace{0.5cm}


\vspace{0.5cm}


\newpage
\section{Introduction}

Information entropies in nonextensive systems
have been extensively investigated since
Tsallis \cite{Tsallis88,Tsallis98} proposed the generalized entropy
(called Tsallis entropy hereafter) defined by
\begin{eqnarray}
S_q &=& \frac{1}{(q-1)}\left[1 - \int p(x,t)^q\:dx \right],
\label{eq:F1}
\end{eqnarray}
where $q$ is the entropic parameter and $p(x,t)$ denotes 
the probability distribution of a state variable $x$ at time $t$.
The Tsallis entropy is not extensive for $q \neq 1.0$
in the sense that the total entropy of a given system
is not proportional to the number of constituent elements.
In the limit of $q=1.0$, $S_q$ reduces to
the Boltzmann-Gibbs-Shannon entropy.
The generalized Fisher information, which is derived for
the generalized Kullback-Leibler divergence 
in conformity with the Tsallis entropy, 
is expressed by \cite{Plastino95}-\cite{Hasegawa08a}
\begin{eqnarray}
g_{ij} &=& 
q\:E\left[ \left( \frac{\partial \ln p(x,t)}
{\partial \theta_i}\right) 
\left( \frac{\partial \ln p(x,t)}
{\partial \theta_j}\right) \right],
\label{eq:A1}
\end{eqnarray}
where  $E[\cdot]$ denotes 
the average over the probability $p(x,t)$ characterized 
by a set of parameters of $\{ \theta_i \}$.
The Tsallis entropy and the generalized Fisher information provide 
the important measure of the information in nonextensive systems.
The Boltzman-Gibbs-Shannon-Tsallis entropy
represents a global measure of ignorance while
the Fisher information expresses a local measure 
of positive amount of information \cite{Frieden98}. 
In particular, the Fisher information signifies the distance between
the neighboring points in the Riemann space spanned
by probability distributions in the information geometry. 
Its inverse expresses the lower bound of the decoding error 
for unbiased estimator in the Cram\'{e}r-Rao inequality.

In a previous paper \cite{Hasegawa08a} (referred to as I hereafter), 
we calculated the Tsallis entropy and generalized Fisher information 
in a microscopic nonextensive system described by the
Langevin model subjected to additive and multiplicative noise.
Employing an analytic time-dependent solution of the
relevant Fokker-Planck equation (FPE), we discussed
the stationary and dynamical properties of the Tsallis entropy $S_q$
and the generalized Fisher information $g_{xx}$ given by
\begin{eqnarray}
g_{xx} &=& q\:E\left[ \left( \frac{\partial \ln p(x,t)}
{\partial x}\right)^2 \right],
\label{eq:A2}
\end{eqnarray}
in stead of $g_{ij}$ given by Eq. (\ref{eq:A1}).
Recently Konno and Watanabe \cite{Konno07} have made 
a detailed study of the Fisher information $g_{ij}$ in 
the stationary state of the Langevin model subjected to cross-correlated 
additive and multiplicative noise, related discussion being 
given in Sec. 4.
It is worthwhile to investigate the dynamics of various elements 
of the generalized Fisher information $g_{ij}$ besides $g_{xx}$.
The purpose of the present paper is to perform detailed calculations
of the time-dependent Tsallis entropy and  generalized Fisher information 
with the use of the analytic $q$-moment approach 
to the FPE which was previously developed in I. 

The paper is organized as follows. In Section 2, we discuss the adopted 
Langevin model and $q$-moment method
to obtain the analytic solution of the FPE.
With the use of the calculated time-dependent distributions,
analytical expressions for the Tsallis entropy and
generalized Fisher information are derived in Section 3. 
In Section 4, we present some model calculations of 
noise-intensity dependences of the generalized Fisher information 
in the stationary state
as well as the dynamical response of
Tsallis entropy and generalized Fisher information
to an applied pulse and sinusoidal inputs. 
Section 5 is devoted to conclusion and discussion on 
relevant previous studies.
\section{Adopted model and method}

\subsection{The Fokker-Planck equation}

We have adopted the Langevin model subjected to additive ($\xi$)
and multiplicative noise ($\eta$) given by
\begin{eqnarray}
\frac{dx}{dt}\!\!&=&\!\! F(x) + G(x) \eta(t)
+ \xi(t)+I(t),
\label{eq:B1} 
\end{eqnarray}
where $F(x)$ and $G(x)$ are arbitrary functions of $x$,
$I(t)$ stands for an external input, and $\eta(t)$ and $\xi(t)$ 
express zero-mean Gaussian white noises with correlations given by
\begin{eqnarray}
\langle \eta(t)\:\eta(t') \rangle
&=& \alpha^2 \:\delta(t-t'),\\
\langle \xi(t)\:\xi(t') \rangle 
&=& \beta^2 \: \delta(t-t'),\\
\langle \eta(t)\:\xi(t') \rangle &=& 0,
\label{eq:B2}
\end{eqnarray}
$\alpha$ and $\beta$ denoting the strengths of multiplicative
and additive noises, respectively.

The FPE in the Stratonovich representation is expressed by
\begin{eqnarray}
\frac{\partial}{\partial t}\: p(x,t) 
&=& - \frac{\partial}{\partial x} \left[\{ F(x) + I(t)\} p(x,t) \right]
+ \left( \frac{\beta^2}{2} \right)
\frac{\partial^2}{\partial x^2}p(x,t) \nonumber
\\
&+& \left( \frac{\alpha^2}{2} \right)
\frac{\partial}{\partial x} 
\left[ G(x) \frac{\partial}{\partial x} \{ G(x) p(x,t) \} \right].
\label{eq:B3}
\end{eqnarray}
Although we have adopted the single-variable Langevin model
in this study, it is straightforward to extend it to the
coupled Langevin model with the use of the mean-field
approximation \cite{Hasegawa08a}.

\subsection{Stationary distribution}

For the linear Langevin model given by
\begin{eqnarray}
F(x) &=& -\lambda x, \label{eq:B15}\\
G(x) &=& x, \label{eq:B16}
\end{eqnarray}
where $\lambda$ denotes the relaxation rate, 
the stationary probability distribution $p(x)$ is expressed 
by \cite{Hasegawa08a}
\begin{eqnarray}
p(x) &=& \frac{1}{Z} 
\left[1- \frac{(1-q)}{2 \phi^2} x^2 \right]^{1/(1-q)}
\exp[Y(x)],
\label{eq:B4}
\end{eqnarray}
with
\begin{eqnarray}
q-1 &=& \frac{2 \alpha^2}{(2\lambda+\alpha^2)}, 
\label{eq:B5}\\
\phi^2 &=&
= \frac{\beta^2}{(2\lambda+\alpha^2)}, 
\label{eq:B6}\\
Y(x)&=& 2 c \:\tan^{-1} \left( \sqrt{\frac{(q-1)}{2\phi^2}}\:x \right), \\
c &=& \frac{I}{\alpha \beta}, \\
Z &=& \left[ \frac{2\phi^2}{\left( q-1 \right)} \right]^{1/2}
\frac{\sqrt{\pi}\:\Gamma(\frac{1}{q-1})\Gamma(\frac{1}{q-1}-\frac{1}{2} )}
{\:\mid \Gamma(\frac{1}{q-1} + i c) \mid^2},
\label{eq:B7}
\end{eqnarray}
where $\Gamma(z)$ denotes the gamma function.
In deriving Eq. (\ref{eq:B7}), 
we have employed the following formula \cite{Konno07}:
\begin{eqnarray}
\int_0^{\pi/2} \cos^u(y)\:\cosh(v y)\:dy
&=& \frac{\pi \Gamma(u+1)}{2^{u+1} \mid \Gamma([u+2+iv]/2) \mid^2},
\label{eq:B8}\\
\Gamma(2z) &=& \frac{2^{2z}}{2 \sqrt{\pi}}
\; \Gamma(z) \Gamma\left(z+\frac{1}{2}  \right).
\label{eq:B9}
\end{eqnarray}
To make the expression of $p(x)$ compact, we have introduced new 
parameters $a$ and $b$, defined by
\begin{eqnarray}
a &=& \left[ \frac{(q-1)}{2 \phi^2} \right]^{1/2}
=\frac{\alpha}{\beta},
\label{eq:B11} \\
b &=& \frac{1}{(q-1)}
=\frac{(2\lambda+\alpha^2)}{2 \alpha^2},
\label{eq:B12} 
\end{eqnarray}
which lead to
\begin{eqnarray}
p(x) &=& \left( \frac{1}{Z} \right)
\frac{\exp\left[2 c \:\tan^{-1} (a x) \right]}
{\left(1+ a^2 x^2 \right)^{b}},
\label{eq:B13}
\end{eqnarray}
with
\begin{eqnarray}
Z &=& \frac{\sqrt{\pi}\:\Gamma(b)\Gamma(b-\frac{1}{2})}
{a\:\mid \Gamma(b+ i c) \mid^2}.
\label{eq:B14}
\end{eqnarray}

We define the $n$th $q$-moment defined by
\begin{eqnarray}
E_q[x^n] &=& \int P_q(x)\;x^n \:dx,
\label{eq:C6} 
\end{eqnarray}
where $E_q[\cdot]$ expresses the average over the escort distribution 
$P_q(x)$ given by
\begin{eqnarray}
P_q(x) &=& \frac{p(x)^q}{c_q}, 
\label{eq:C7}\\
c_q &=& \int p(x)^q \: dx,
\label{eq:C8}
\end{eqnarray}
while $E[\cdot]$ denotes the average over $p(x)$ 
[{\it e.g.} Eq. (\ref{eq:A1})].
By using Eqs. (\ref{eq:C6})-(\ref{eq:C8})
for the stationary distribution given by Eq. (\ref{eq:B13}),
we may obtain 
the variance $\mu_q$ ($=E_q[x]$) and 
covariance $\sigma_q^2$ ($=E_q[x^2]-E_q[x]^2$)
in the stationary state given by
\begin{eqnarray}
\mu_q &=& \frac{c(b+1)}{a b^2}
=\frac{2(2 \lambda+3\alpha^2)I}{(2\lambda+\alpha^2)^2}
\simeq \frac{I}{\lambda}, 
\label{eq:C9} \\
\sigma_q^2 &=& \frac{[b^2+(qc)^2]}{a^2b^2(2b-1)}
= \frac{(\alpha^2 \mu_q^2+\beta^2)}{2 \lambda}.
\label{eq:C10}
\end{eqnarray}

Depending on the model parameters, the stationary distribution 
given by Eq. (\ref{eq:B4}) or (\ref{eq:B13}) may reproduce various 
distributions such as the Gaussian ($q=1.0$), $q$-Gaussian ($q \neq 1.0$, 
$c = 0.0$), Cauchy ($q=2.0$, $c \neq 0.0$) and inverse-gamma distributions 
($\beta=0.0$, $c \neq 0.0$) \cite{Hasegawa08a}.

\subsection{Dynamical distribution}

In order to obtain the dynamical solution of the FPE given by 
Eq. (\ref{eq:B3}), we have adopted the following $q$-moment
approach \cite{Hasegawa08a}.

(1) From an equation of motion for the $n$th $q$-moment of $E_q[x^n]$ 
given by Eq. (\ref{eq:C6}), 
\begin{eqnarray}
\frac{d E_q[x^n]}{d t} &=& \frac{d}{d t}
\int P_q(x,t)\:x^n \:dx,\\
&=& \frac{q}{c_q}\int 
\left( \frac{\partial p(x,t)}{\partial t}\right)\:p(x,t)^{q-1}\:x^n \:dx
-\frac{1}{c_q}\left( \frac{d c_q}{dt} \right) E_q[x^n],
\\
\frac{d c_q}{dt} 
&=& q \int \left( \frac{\partial p(x,t)}{\partial t}\right)\:p(x,t)^{q-1}\:dx,
\end{eqnarray} 
we have derived equations of motion for $\mu_q(t)$ and $\sigma_q(t)^2$
given by \cite{Hasegawa08a}
\begin{eqnarray}
\frac{d \mu_q(t)}{dt} &\simeq& -\lambda \mu_q(t) 
+ I(t),
\label{eq:C4} \\
\frac{d \sigma_q(t)^2}{dt} &\simeq& -2 \lambda \sigma_q(t)^2 
+ \alpha^2 \mu_q(t)^2+\beta^2.
\label{eq:C5}
\end{eqnarray}

(2) We have assumed that the dynamical distribution 
has the same functional form as that of the stationary one, 
\begin{eqnarray}
p(x,t) &=& \frac{1}{Z(t)}
\frac{\exp[2c(t) \tan^{-1}a(t) x]}{[1+a(t)^2 x^2]^{b}},
\label{eq:C1}
\end{eqnarray}
with the time-independent $b$ [$=1/(q-1)$]
and the time-depedent $a(t)$ and $c(t)$ given by
\cite{Hasegawa08a}
\begin{eqnarray}
a(t) &=& \frac{1}{[(2b-1)\sigma_q(t)^2-\mu_q(t)^2]^{1/2}},
\label{eq:C2} \\
c(t) &=& \frac{a(2b-1) \mu_q(t)}{2}
\label{eq:C3} \\
Z(t) &=& \frac{\sqrt{\pi}\:\Gamma(b)\Gamma(b-\frac{1}{2})}
{a(t)\:\mid \Gamma[b+ i c(t)] \mid^2}.
\label{eq:C0}
\end{eqnarray}
where $\mu_q(t)$ and $\sigma_q(t)^2$ obey equations of
motion given by Eqs. (\ref{eq:C4}) and (\ref{eq:C5}).
Equations (\ref{eq:C2}) and (\ref{eq:C3}) are derived from 
the relations for the stationary state given 
by Eqs. (\ref{eq:C9}) and (\ref{eq:C10}) after I 
(see Eqs. (86)-(90) in Ref. \cite{Hasegawa08a}).

With the assumption (2), we once
tried to derive equations of motion for
$a(t)$, $b(t)$, $c(t)$ and $Z(t)$ from the FPE, but failed:
they are not determined in an appropriate way \cite{Hasegawa08a}.
Then we have decided to express the parameters of $a(t)$ and $c(t)$
in terms of $\mu_q(t)$ and $\sigma_q(t)^2$,
as given by Eqs.  (\ref{eq:C2}) and (\ref{eq:C3}).
The time-dependent distributions calculated 
by Eqs. (\ref{eq:C4})-(\ref{eq:C0}) are in good agreement 
with those obtained by the partial difference equation method (PDE),
as was discussed in I (see Fig. 7 of Ref. \cite{Hasegawa08a}) 
and will be shown shortly (Fig. \ref{figG}).

\section{Information entropies}
\subsection{Tsallis Entropy}

In a previous study \cite{Hasegawa08a}, we numerically calculated 
the Tsallis entropy $S_q$ given by Eq. (\ref{eq:F1}).
By using the time-dependent distribution $p(x,t)$
given by Eq. (\ref{eq:C1}), 
we may obtain analytical expression for $S_q$ given by
\begin{eqnarray}
S_q &=& \frac{1}{2} [1+\ln (2 \pi \sigma_q^2)] 
\hspace{1cm} \mbox{(for $q=1$)}, \\
&=& \left( \frac{1-c_q}{q-1} \right),
\hspace{2cm} \mbox{(for $q \neq 1$)},
\label{eq:F7}
\end{eqnarray}
with
\begin{eqnarray}
c_q &=& Z^{1-q} \left( \frac{b(2b-1)}{2[b^2+(qc)^2]} \right)
\frac{\vert \Gamma(b+ic) \vert^2}{\vert \Gamma(b+iqc) \vert^2},
\label{eq:F3}
\end{eqnarray}
where $Z$ is given by Eq. (\ref{eq:C0}).

\subsection{Generalized Fisher information}

In order to derive the analytic expression for
the generalized Fisher information, we first calculate 
the derivatives of $ \ln p(x,t)$
with respect to ${\sf \theta}=(a, b, c)$, as given by
(the argument $t$ in $a$ and $b$ being suppressed),
\begin{eqnarray}
\frac{\partial \ln p(x,t)}{\partial a} 
&=& 2c \left(\frac{x}{U}\right)
+\frac{2b}{a} \left(\frac{1}{U}\right) 
-\frac{(2b-1)}{a}, 
\label{eq:D2}\\
\frac{\partial \ln p(x,t)}{\partial b}
&=& -\ln U+A, 
\label{eq:D3} \\
\frac{\partial \ln p(x,t)}{\partial c}
&=& 2 \tan^{-1}(a x)- B, \label{eq:D4} 
\end{eqnarray}
with
\begin{eqnarray}
U &=& 1+a x^2, \label{eq:D5} \\
A&=& 2 {\rm Re} \:\psi(b+ic)-\psi(b)-\psi(b-1/2), 
\label{eq:D6} \\
B&=& 2 {\rm Im} \:\psi(b+ic),
\label{eq:D7}
\end{eqnarray}
where the relation given by $x^2/U=(1-1/U)/a^2$ is employed.
Substituting Eqs. (\ref{eq:D2})-(\ref{eq:D4}) to Eq. (\ref{eq:A1}) 
and evaluating various averages such as $E[1/U]$ and 
$E[\ln U]$ after the method mentioned in the Appendix A
(and the Table 1),
we have obtained the generalized Fisher information matrix $G$,
whose elements $g_{ij}$ [$=(G)_{ij}$] are expressed by
\begin{eqnarray}
g_{aa} &=& \frac{q (2b-1)(b+1+c^2)}
{a^2[(b+1)^2+c^2]}, 
\label{eq:D8} \\
g_{bb} &=& q\:[\psi'(b)+\psi'(b-1/2)-2 {\rm Re} \:\psi'(b+ic)], 
\label{eq:D9} \\
g_{cc} &=& 2q\: {\rm Re}\: \psi'(b+ic), 
\label{eq:D10} \\
g_{ab} &=& g_{ba} = \frac{q(b+2c^2)}{a(b^2+c^2)},
\label{eq:D11} \\
g_{ac} &=& g_{ca} = - \frac{qc(2b-1)}{a(b^2+c^2)},
\label{eq:D12}\\
g_{bc} &=& g_{cb} = 2q \:{\rm Im}\: \psi'(b+ic).
\label{eq:D13}
\end{eqnarray}
In the limit of $c=0$, Eqs. (\ref{eq:D8})-(\ref{eq:D13}) reduce to
\begin{eqnarray}
g_{aa} &=& \frac{q(2b-1)}{a^2 (b+1)}, 
\label{eq:D14}\\
g_{bb} &=& q[\psi'(b-1/2)-\psi'(b)],
\label{eq:D15} \\
g_{cc} &=& 2q \:\psi'(b),
\label{eq:D16} \\
g_{ab} &=& g_{ba} = \frac{q}{ab},
\label{eq:D17} 
%
\end{eqnarray}
and $g_{ij}=0$ otherwise.

\section{Model calculations}
\subsection{Stationary properties}

By using Eqs. (\ref{eq:D8})-(\ref{eq:D13}),
we have performed some model calculations of 
the generalized Fisher information matrix $G$, 
whose dependences on $\alpha$, $\beta$ and $I$
are plotted in Fig. \ref{figH}(a)-(f):
relevant calculations of the Tsallis entropy were presented
in I (see Figs. 2, 3 and 4 in Ref. \cite{Hasegawa08a}).

\noindent
{\it (a) $\alpha$ dependence}

First we show in Figs. \ref{figH}(a) and (b),
the diagonal and off-diagonal elements of $G$
as a function of $\alpha^2$ for $\beta=0.5$ and $I=1.0$.
We note that with increasing $\alpha$, $g_{aa}$ is significantly
decreased while $g_{bb}$ and $g_{cc}$ are increased.
$\vert g_{bc} \vert$ is gradually increased with a negative sign 
whereas the saturation behavior is realized in $g_{ab}$ and $g_{ac}$
for a large $\alpha$. 

\noindent
{\it (b) $\beta$ dependence}

Figures \ref{figH}(c) and (d) show the $\beta$ dependence
of the diagonal and off-diagonal elements, respectively,
of $G$ for $\alpha=0.5$ and $I=1.0$.
With increasing $\beta$, $g_{aa}$ and $g_{cc}$ are increased
whereas $g_{bb}$ is decreased.
Off-diagonal elements are small at $\beta=0.0$.
With increasing $\beta$ from zero value,
$g_{ab}$ is increased with a positive sign
while $ \vert g_{bc} \vert $ and $ \vert g_{ac} \vert $ are increased
with negative signs.

\noindent
{\it (c) $I$ dependence}

The $I$ dependences of diagonal and off-diagonal elements
of $G$ are plotted in
Figs. \ref{figH}(e) and (f), respectively.
With increasing $I$,
both $g_{aa}$ and $g_{bb}$ are increased whereas $g_{cc}$
is decreased.
Although off-diagonal elements are vanishing at $I=0.0$ 
except for $g_{ab}$ as Eq. (\ref{eq:D17}) shows,
a positive $g_{ab}$ is increased while $ \vert g_{bc} \vert$
and $ \vert g_{ac} \vert $ are increased with negative signs
when $I$ is increased. 

\subsection{Dynamical properties}

We will discuss the response of the system
to applied external signals in this subsection.
In order to examine the validity of the analytic dynamical approach, 
we have employed the PDE derived from Eq. (\ref{eq:B3}), 
as given by
\begin{eqnarray}
p(x,t+v) &=& p(x,t)
+\left(-F'(x)
+ \frac{\alpha^2}{2}[G'(x)^2+G(x)G^{(2)}(x)] \right) v \:p(x,t)
\nonumber \\
&+&\left[ -F(x)-I(t)
+ \frac{3 \alpha^2}{2}  G(x)G'(x) \right]
\left(\frac{v}{2 u}\right)[p(x+u)-p(x-u)] \nonumber \\
&+& \left(\frac{\alpha^2}{2} G(x)^2 + \frac{\beta^2}{2} \right)
\left(\frac{v}{u^2}\right)[p(x+u,t)+p(x-u,t)-2p(x,t)],
\label{eq:M1}
\end{eqnarray}
where $u$ and $v$ denote incremental steps of $x$ and $t$, respectively.
We impose the boundary condition:
\begin{eqnarray}
p(x,t)=0, \hspace{1cm}\mbox{for $ \mid x \mid \ge x_m$}
\label{eq:M2}
\end{eqnarray}
with $x_m=5$, and the initial condition of $p(x,0)=p_0(x)$ where $p_0(x)$ 
is the stationary distribution given by Eqs. (\ref{eq:B13}) and (\ref{eq:B14}).
We have chosen parameters of $u=0.05$ and $v=0.0001$ such as to satisfy 
the condition: $(\alpha^2 x_m^2 v/2 u^2) < 1/2$, which is required for
stable, convergent solutions of the PDE \cite{Hasegawa08a}.

\vspace{0.5cm}
\noindent
{\it (a) Pulse Input}

We first apply a pulse input given by
\begin{eqnarray}
I(t) &=& \Delta I
\: [\Theta(t-2)\Theta(6-t)-\Theta(t-10)\Theta(14-t)],
\label{eq:E1}
\end{eqnarray}
to a system with $\lambda=1.0$, $\alpha=0.5$ and $\beta=0.5$,
where $ \Delta I=1.0$ and $\Theta(x)$ denotes the Heaviside function:
$\Theta(x)=1$ for $x \geq 1$ and zero for $x < 0$.
Positive and negative pulses with the magnitude 
of $\vert \Delta I \vert$
are applied at $2.0 \leq t < 6.0$ and $10.0 \leq t < 14.0$,
respectively.
The time-dependent distributions of
$p(x,t)$ at $0 \leq t < 10.0$ are plotted in
Fig. \ref{figG}, where solid (dashed) curves express
the results of the $q$-moment (PDE) method \cite{Hasegawa08a}.
By a positive pulse applied at $2.0 \leq t < 6.0$, $p(x,t)$ moves rightward
with a slight modification of its shape. In contrast,
when a negative pulse is applied at $10.0 \leq t < 14.0$,
$p(x,t)$  moves leftward again with a slight modification of its shape.
We note that the results calculated by the $q$-moment method 
are in good agreement with those obtained by the PDE method.

Figure \ref{figA}(a) shows the time dependence of $\mu_q(t)$ and 
$\sigma_q(t)^2$ in response to the applied pulse input.
We note that $\mu_q$ is increased (decreased)
by an applied positive (negative) pulse  
whereas $\sigma_q^2$ is increased for both positive and negative
inputs. The result calculated with equations of motion given by
Eqs. (\ref{eq:C4}) and (\ref{eq:C5}) shown by solid curves well agrees
with those obtained by the PDE method shown by dashed curves.

The Tsallis entropy $S_q(t)$ 
in response to an applied pulse is plotted in
Fig. \ref{figA}(b), where the solid and dashed curves
express the results calculated 
by Eqs. (\ref{eq:F7}) and (\ref{eq:F3})
and by the PDE method, respectively.
It is shown that $S_q(t)$ is increased for both positive 
and negative input pulses.

The time dependences of the diagonal and off-diagonal elements of 
the generalized Fisher information are plotted in Figs. \ref{figB}(a) and 
\ref{figB}(b), respectively. When a positive input signal is applied 
at $2.0 \leq t < 6.0$, $g_{aa}$, $g_{bb}$ and $g_{ab}$ are increased
whereas $g_{cc}$, $g_{bc}$ and $g_{ac}$ are decreased.
When a negative input signal is applied at $10.0 \leq t < 14.0$, 
all the elements except for $g_{ac}$ show the same behaviors as those 
for a positive input: only $g_{ac}$ is expressed by an odd function of $c$ 
in Eqs. (\ref{eq:D8})-(\ref{eq:D13}).

Figure \ref{figC} shows the time dependence of inverse of 
the generalized Fisher information matrix, $h_{ij}$ [$=(G^{-1})_{ij}$]. 
With an applied pulse, $h_{cc}$ is increased while $h_{aa}$ 
and $h_{bb}$ are decreased.
It implies from the Cram\'{e}r-Rao theorem that 
when an external signal is applied, 
the lower bound in estimating the parameter $c$ is increased while 
those in estimating $a$ and $b$ are decreased.

\vspace{0.5cm}
\noindent
{\it (b) Sinusoidal Input}

Next we apply a sinusoidal input given by
\begin{eqnarray}
I(t) &=& \Delta I\: \sin \left( \frac{2 \pi (t-2)}{T_p} \right)
\:\Theta(t-2)\Theta(T_p+2-t),
\label{eq:E2}
\end{eqnarray}
where $ \Delta I=1.0$ and $T_p=10.0$. 
The time dependences of $\mu_q(t)$ and 
$\sigma_q(t)^2$ in response to the applied sinusoidal input
are plotted in Figure \ref{figF}(a), where 
bold and thin solid curves express the results calculated
by Eqs. (\ref{eq:C4}) and (\ref{eq:C5}), and where dashed curves 
show those obtained by the FPE method.
The response of $\mu_q(t)$ lags behind the input $I(t)$
by about 0.85.
The time dependence of $S_q(t)$ is plotted in Fig. \ref{figF}(b),
where solid and dashed curves denote the results
of the $q$-moment and PDE methods, respectively.
Figure \ref{figE}(a) and (b) show
the time dependences of diagonal and off-diagonal elements,
respectively, of the generalized Fisher information.
A comparison of Figs. \ref{figF} and \ref{figE}
to Figs. \ref{figA} and \ref{figB} shows
that time dependences of $\mu(t)$, $\sigma_q(t)^2$,
$S_q(t)$ and $g_{ij}(t)$ for a sinusoidal input
are not dissimilar to those for a pulse input.  
This is true also for the time dependence of the inversed
Fisher information (relevant results not shown).

\section{Discussion and conclusion}

In a previous study \cite{Hasegawa08a}, we numerically calculated 
the generalized Fisher information  $g_{xx}$ given by Eq. (\ref{eq:A2}).
By using the dynamical distribution given by Eq. (\ref{eq:C1}), 
we may analytically calculate $g_{xx}$ given by
\begin{eqnarray}
g_{xx} &=& \frac{q a^2b(b+1)(2b-1)}{[(b+1)^2+c^2]},
\label{eq:F2} \\
&=& \frac{q \:a^2 b(2b-1)}{(b+1)}
= \frac{(3-q)}{2 \phi^2}
= \frac{1}{\sigma_q^2}.
\label{eq:F4} \hspace{1cm} \mbox{(for $c=0.0$)}.
\end{eqnarray}
The time dependence of $g_{xx}$ when a pulse input given 
by Eq. (\ref{eq:E1}) is applied, is plotted by the chain curve 
in Fig. \ref{figD}(a).

For a comparison, we have calculated the generalized Fisher information 
for $\theta_i=\theta_j=\mu_q$ and 
$\theta_i=\theta_j=\sigma_q^2 \equiv \gamma$ in Eq. (\ref{eq:A1}), 
by using the relations given by
\begin{eqnarray}
g_{\mu\mu} &=& \left( \frac{\partial a}{\partial \mu_q} \right)^2 g_{aa}
+ \left( \frac{\partial c}{\partial \mu_q} \right)^2 g_{cc}
+ 2 \left( \frac{\partial a}{\partial \mu_q} \right)
\left( \frac{\partial c}{\partial \mu_q} \right)g_{ac}, 
\label{eq:L1} \\
g_{\gamma\gamma} 
&=& \left( \frac{\partial a}{\partial \sigma_q^2} \right)^2 g_{aa},
\label{eq:L2}
\end{eqnarray}
where $\partial a/\partial \mu_q$ {\it et. al.} may be
calculated from Eqs. (\ref{eq:C2}) and (\ref{eq:C3}).
Solid curves in Figs. \ref{figD}(a) and (b) show
the time dependences of
$g_{\mu\mu}$ and $g_{\gamma \gamma}$, respectively,
in response of a pulse input given by Eq. (\ref{eq:E1}).
When an input pulse is applied, $g_{\mu\mu}$
is decreased while $g_{\gamma\gamma}$ is increased.
We note that $g_{xx}$ and $g_{\mu \mu}$
are almost the same in the stationary state
and that their responses to an input
are similar apart from their magnitudes.

It is well known that the relation given by
\begin{eqnarray}
g_{\mu \mu}& = & g_{xx},
\label{eq:F5}
\end{eqnarray}
holds for $q=1.0$ ($\alpha=0.0$) 
because the time-dependent Gaussian distribution 
is given by 
\begin{equation}
p(x,t)= \frac{1}{\sqrt{2 \pi \:\sigma(t)^2}}
\;e^{-[x-\mu(t)]^2/2 \sigma(t)^2},
\label{eq:K1}
\end{equation}
with $\mu(t)$ and $\sigma(t)^2$ satisfying equations of motion given by
\begin{eqnarray}
\frac{d \mu(t)}{dt} &=& -\lambda \mu(t)+ I(t), 
\label{eq:K2} 
\\
\frac{d \sigma(t)^2}{dt} &=& -2 \lambda \sigma(t)^2 + \beta^2.
\label{eq:K3}
\end{eqnarray}
The relation given by Eq. (\ref{eq:F5}) is obtainable also
for the $q$-Gaussian distribution given by
\begin{eqnarray}
p(x,t) &=& \frac{1}{Z_q}\left[1-\frac{(1-q)}{(3-q) \sigma_q(t)^2}
\{ x-\mu_q(t) \}^2 \right]^{1/(1-q)},
\label{eq:K4}
\end{eqnarray}
which is derived from the maximum-entropy method
for given $\mu_q$ and $\sigma_q^2$.
It is natural that Figs. \ref{figD}(a) shows
$g_{\mu \mu} \neq g_{xx}$ for $I(t) \neq 0.0$ because 
the distribution given by Eq. (\ref{eq:C1}) 
is rather different from the $q$-Gaussian given by Eq. (\ref{eq:K4})
except for $c = \mu_q = 0.0$ 
(see Figs. 10 and 11 of Ref. \cite{Hasegawa08a}).

Although additive and multiplicative noise is assumed to be uncorrelated 
in Sec. 2, we may take into account effects of the cross-correlation between 
the two noise, introducing the degree of the cross-correlation 
$\epsilon$ with 
$\langle \eta(t)\:\xi(t') \rangle = \epsilon \alpha \beta \:\delta(t-t')$
in place of Eq. (\ref{eq:B2}).
The stationary distribution is given by (for details, see the Appendix B)
\begin{eqnarray}
p(x) &=& \left( \frac{1}{Z} \right)
\frac{\exp\left[2 c \:\tan^{-1} a (x+f) \right]}
{\left[1+ a^2 (x+f)^2 \right]^{b}},
\label{eq:G1}
\end{eqnarray}
with
\begin{eqnarray}
a &=& \frac{\alpha}{\beta \sqrt{1-\epsilon^2}}, \\
b &=& \frac{(2 \lambda + \alpha^2)}{2 \alpha^2}, \\
c &=& \frac{(I+\lambda f)}{\beta \sqrt{1-\epsilon^2}}, \\
f &=& \frac{\epsilon \beta}{\alpha}, \\
Z &=& \frac{\sqrt{\pi}\:\Gamma(b)\Gamma(b-1/2)}
{a\:\mid \Gamma(b+ i c) \mid^2}.
\label{eq:G2}
\end{eqnarray}
The distribution given by Eq. (\ref{eq:G1}),
which is shifted by an amount of $f$ by
an introduced cross-correlation, 
has the same dependence on 
the parameters $a$, $b$ and $c$ as that of Eq. (\ref{eq:B13}).
The Langevin model with the cross-correlated noise
($\epsilon \neq 0.0$) but no external inputs ($I=0.0$) was 
discussed by Konno and Watanabe, who derived useful 
expressions for the generalized Fisher information in its stationary state
\cite{Konno07}.
\footnote{
The factor corresponding to $\epsilon \:\alpha\beta G'(x)$
in the first term of Eq. (\ref{eq:Y1}) is missing 
in Eq. (23) of Konno and Watanabe \cite{Konno07}:
$K(x)=(\alpha-D_p)x$ and $f=2(b-1)c/A$
in Eqs. (23) and (25) of Ref. \cite{Konno07}
should be replaced by $K(x)=(\alpha-D_p)x-D_{ap}$
and $f=(2b-1)c/A$, respectively. 
The probability distribution given by Eq. (27) of Ref. \cite{Konno07}
is valid with these replacements.
}
 
Our result presented in this study may be applied to
another type of the Langevin model, for example, 
with $\epsilon = 0.0$, and $F(x)$ and $G(x)$ given by
\begin{eqnarray}
F(x) &=& -\lambda (x+s), \\
G(x) &=& \sqrt{r^2+ 2 sx+ x^2},
\end{eqnarray}  
where $\lambda $ expresses the relaxation rate,
and $r$ and $s$ are additional parameters.
In the case of $[\beta^2+\alpha^2(r^2-s^2)] > 0$,
the stationary distribution is given by \cite{Hasegawa08c}
\begin{eqnarray}
p(x) &=& \left( \frac{1}{Z} \right)
\frac{\exp\left[2 c \:\tan^{-1} a (x+s) \right]}
{\left[1+ a^2 (x+s)^2 \right]^{b}},
\label{eq:G3}
\end{eqnarray}
with
\begin{eqnarray}
a &=& \frac{\alpha}{\sqrt{\beta^2+\alpha^2(r^2-s^2)}}, \\
b &=& \frac{(2 \lambda + \alpha^2)}{2 \alpha^2}, \\
c &=& \frac{I}{\alpha \sqrt{\beta^2+\alpha^2(r^2-s^2)} }, \\
Z &=& \frac{\sqrt{\pi}\:\Gamma(b)\Gamma(b-\frac{1}{2})}
{a\:\mid \Gamma(b+ i c) \mid^2}.
\end{eqnarray} 
Equation (\ref{eq:G3}) is equivalent to 
Eqs. (\ref{eq:B13}) and (\ref{eq:G1}).
Discussion on the information entropies presented in Sec. 3 
may be applied also to this model. 

Quite recently we have investigated the effect of spatially correlated
variability on the information entropies in nonextensive systems, 
by using the maximum-entropy method \cite{Hasegawa08b}.
It has been shown that effects of the spatially correlated variability 
on the generalize Fisher information
are different from those on the Tsallis entropy:
the generalized Fisher information is increased (decreased) 
by a positive (negative) spatial correlation,
whereas the Tsallis entropy is decreased with increasing 
an absolute magnitude of the correlation independently of its sign.
This fact arises from the difference in their characteristics
\cite{Frieden98}. In Ref. \cite{Hasegawa08b}, we obtained
analytic expressions for the Tsallis entropy
and the generalized Fisher entropy 
in spatially correlated nonextensive systems
from the $q$-Gaussian-type distribution derived by the maximum-entropy method
\cite{Hasegawa08b}. 
Unfortunately, we cannot obtain analytic distributions 
in the Langevin model subjected to spatially correlated
additive and multiplicative noise,
because we have no analytic approaches to
solve the relevant FPE, even for $N=2$ \cite{Hasegawa08c}.
It is necessary to develop an appropriate analytic 
method to solve the FPE including the spatial correlation in 
additive and multiplcative noise. 

To summarize, we have discussed the stationary and
dynamical properties of Tsallis entropy and generalized 
Fisher entropy in the Langevin model subjected to
additive and multiplicative noise, 
which is a typical microscopic nonextensive model.
By employing the dynamical solution of the FPE, we have derived
analytical expressions for the time-dependent information entropies, 
with which model calculations of their responses
to applied pulse and sinusoidal inputs have been performed.
The analytic $q$-moment approach to the FPE developed in I
is useful and applicable to the Langevin model which
includes the time-dependent model parameters \cite{Hasegawa08a}.
It is possible to generalize the moment method to the FPE for various
types of Langevin models, which will be reported in a separate paper 
\cite{Hasegawa08d}.

\section*{Acknowledgements}
This work is partly supported by
a Grant-in-Aid for Scientific Research from the Japanese 
Ministry of Education, Culture, Sports, Science and Technology.  

\appendix

\vspace{0.5cm}
\noindent
{\large\bf Appendix A: Evaluations of averaged values}
\renewcommand{\theequation}{A\arabic{equation}}
\setcounter{equation}{0}

Various quantities averaged over the distribution $p(x)$,
which appear in the Fisher information,
may be evaluated with the use of $K_m$ defined by
\begin{eqnarray}
K_m & \equiv & K_m(a,b,c)= \int_{-\infty}^{\infty}
\frac{e^{2c \tan^{-1}(ax)}}{U^{b+m}}\:dx, 
\label{eq:Z1} \\
&=& \frac{\sqrt{\pi} \:\Gamma(b+m)\Gamma(b+m-1/2)}
{a\:\vert \Gamma(b+m+ic) \vert^2},
\label{eq:Z2}
\end{eqnarray}
where $U=1+a x^2$ and  $K_0 = Z$.
Taking derivatives of Eq. (\ref{eq:Z1}) with respect to
parameters of $a$, $b$ and $c$,
we obtain
\begin{eqnarray}
\frac{K_m}{K_0} &=& E \left[ \frac{1}{U^m} \right],
\label{eq:Z3} \\
\frac{1}{K_0}\frac{\partial K_m}{\partial a}
&=&-2a(b+m) E\left[ \frac{x^2}{U^{m+1} } \right]
+ 2c E\left[ \frac{x}{U^{m+1} } \right],\\
\frac{1}{K_0}\frac{\partial K_m}{\partial b}
&=& - E\left[ \frac{\ln U}{U^m} \right],\\
\frac{1}{K_0}\frac{\partial K_m}{\partial c}
&=& 2 E\left[ \frac{\tan^{-1}(a x)}{U^m} \right],\\
\frac{1}{K_0}\frac{\partial^2 K_m}{\partial a^2}
&=& 4a^2(b+m)(b+m+1) E\left[ \frac{x^4}{U^{m+2}} \right]
-4ac(b+m+1) E\left[ \frac{x^3}{U^{m+2}} \right] \nonumber \\
&-& 4ac(b+m) E\left[ \frac{x^3}{U^{m+1}} \right]
+4c^2 E\left[ \frac{x^2}{U^{m+2}} \right]
-2(b+m) E\left[ \frac{x^2}{U^{m+1}} \right], \\
\frac{1}{K_0}\frac{\partial^2 K_m}{\partial b^2}
&=& E\left[ \frac{(\ln U)^2}{U^m} \right],\\
\frac{1}{K_0}\frac{\partial^2 K_m}{\partial c^2}
&=& 4 E\left[ \frac{[\tan^{-1}(a x)]^2}{U^m} \right],\\
\frac{1}{K_0}\frac{\partial^2 K_m}{\partial a \partial b}
&=& -2a E\left[ \frac{x^2}{U^{m+1} } \right]
+ 2a (b+m) E\left[ \frac{x^2 \ln U}{U^{m+1} } \right]
-2c E\left[ \frac{x \ln U}{U^{m+1} } \right],\\
\frac{1}{K_0}\frac{\partial^2 K_m}{\partial b \partial c}
&=& -2 E\left[ \frac{\ln U \tan^{-1}(ax)}{U^{m} } \right],\\
\frac{1}{K_0}\frac{\partial^2 K_m}{\partial a \partial c}
&=& -4a(b+m) E\left[ \frac{x^2 \tan^{-1}(ax)}{U^{m+1} } \right]
+ 2 E\left[ \frac{x}{U^{m+1} } \right] \nonumber \\
&+& 4c E\left[ \frac{x \tan^{-1}(ax)}{U^{m+1} } \right].
\end{eqnarray}
From derivatives of Eq. (\ref{eq:Z2}) with respect to $a$, $b$ and $c$, 
we obtain 
\begin{eqnarray}
\frac{\partial \ln K_m}{\partial a}
&=& -\frac{1}{a}, \\
\frac{\partial \ln K_m}{\partial b}
&=& \psi(b+m)+\psi(b+m-1/2)-2\:{\rm Re}\:\psi(b+m+ic), \\
\frac{\partial \ln K_m}{\partial c}
&=& 2 \:{\rm Im} \:\psi(b+m+ic), \\
\frac{\partial^2 \ln K_m}{\partial a^2}
&=& \frac{1}{a^2}, \\
\frac{\partial^2 \ln K_m}{\partial b^2}
&=& \psi'(b+m)+\psi'(b+m-1/2)-2\:{\rm Re}\:\psi'(b+m+ic), \\
\frac{\partial^2 \ln K_m}{\partial c^2}
&=& 2 \:{\rm Re} \:\psi'(b+m+ic), \\
\frac{\partial^2 \ln K_m}{\partial b\partial c}
&=& 2 \:{\rm Im} \:\psi'(b+m+ic), \\
\frac{\partial^2 \ln K_m}{\partial a\partial b}
&=& \frac{\partial^2 K_m}{\partial a\partial c}=0.
\label{eq:Z4}
\end{eqnarray}
By using Eqs. (\ref{eq:Z3})-(\ref{eq:Z4}) and the relations given by
\begin{eqnarray}
\frac{x^2}{U} &=& \frac{1}{a^2}\left( 1-\frac{1}{U} \right), \\
\frac{1}{K_0}\frac{\partial^2 K_m}{\partial a \partial b} 
&=& \frac{K_m}{K_0} \left[
\left( \frac{\partial \ln K_m}{\partial a} \right)
\left( \frac{\partial \ln K_m}{\partial b} \right)
+ \left( \frac{\partial^2 \ln K_m}{\partial a \partial b} \right)
\right],
\end{eqnarray}
we may calculate averaged values of various quantities,
whose results are summarized in the Table 1.

\vspace{0.5cm}
\noindent
{\large\bf Appendix B: The Langevin model with cross-correlated
noise}
\renewcommand{\theequation}{B\arabic{equation}}
\setcounter{equation}{0}

We assume the Langevin model subjected to cross-correlated additive
and multiplicative noise, which is
given by Eqs. (\ref{eq:B1})-(\ref{eq:B2}) 
but Eq. (\ref{eq:B2}) is replaced by
\begin{eqnarray}
\langle \eta(t)\:\xi(t') \rangle 
&=& \epsilon \:\alpha \beta \: \delta(t-t'),
\end{eqnarray}
with $\epsilon$ expressing the degree of the cross-correlation. 

The FPE for the Langevin model is given by \cite{Tessone98,Liang04,Jin05}
\begin{eqnarray}
\frac{\partial}{\partial t}\: p( x,t) 
&=&- \frac{\partial}{\partial x}\left( \left[F(x) +I
+\left( \frac{\phi}{2} \right)
[\alpha^2G'(x)G(x)+ \epsilon \alpha \beta\: G'(x)]
\right] \:p( x,t)\right)  
\nonumber \\
&+& \left(\frac{1}{2} \right) \frac{\partial^2}{\partial x^2} 
\{[\alpha^2 G(x)^2+ 2 \epsilon \alpha\beta G(x)+\beta^2]\:p(x,t) \},
\label{eq:Y1}
\end{eqnarray}
where $\phi=1$ and 0 in the Stratonovich and
Ito representations, respectively.

The stationary probability distribution $p(x)$
is expressed by 
\begin{eqnarray}
\ln p(x) &=& 2 \int  \:
\left[ \frac{F(x)+I}
{\alpha^2 G(x)^2+2\epsilon\alpha\beta G(x)+\beta^2} \right]\:dx \nonumber\\
&-&\left(1- \frac{\phi}{2} \right)
\ln \left( \frac{1}{2} \:[\alpha^2 G(x)^2+ 2 \epsilon \alpha\beta G(x)
+\beta^2 ]\right).
\label{eq:Y2}
\end{eqnarray}
For the linear Langevin model given by
$F(x)=-\lambda x$ and $G(x)=x$, Eq. (\ref{eq:Y2}) 
in the Stratonovich representation ($\phi=1$)
yields the stationary distribution
expressed by Eqs. (\ref{eq:G1})-(\ref{eq:G2}).

\newpage

\vspace{1cm}
{\it Table 1} 
Averaged values of various quantities:
$E[Q(x)]=\int p(x) \:Q(x) \:dx\;\;$ 

\begin{center}
\begin{tabular}{|c|c|}
\hline
$ Q(x) $ & $E[ Q(x) ]$ \\ \hline \hline
$x$ & $c/a(b-1)$ \\
$x^2$ & $(b-1+2c^2)/a^2(b-1)(2b-3)$ \\
$x^2-(E[x])^2$ & $[(b-1)^2+c^2]/a^2(b-1)^2(2b-3)$ \\
$1/U $ & $b(2b-1)/2(b^2+c^2)$ \\
$1/U^2$ & $b(b+1)(2b+1)(2b-1)/4(b^2+c^2)[(b+1)^2+c^2]$ \\
$x/U$ & $c(2b-1)/2a(b^2+c^2)$ \\
$x/U^2$ & $bc(2b+1)(2b-1)/4a(b^2+c^2)[(b+1)^2+c^2]$ \\
$\ln U$ & $ -[\psi(b)+\psi(b-1/2)-2 {\rm Re}\:\psi(b+ic)]$ \\
$(\ln U)^2 - (E[\ln U])^2$ & 
$ \psi'(b)+\psi'(b-1/2)-2 {\rm Re}\:\psi'(b+ic)$\\
$\tan^{-1}(ax)$ & ${\rm Im} \:\psi(b+ic)$ \\
$[\tan^{-1}(ax)]^2 - (E[\tan^{-1}(ax)])^2 
$ & ${\rm Re} \:\psi'(b+ic)/2 $ \\
\hline
\end{tabular}
\end{center}

\noindent
\hspace{2cm}($U=1+a^2 x^2$)

\newpage


\newpage

\begin{figure}
\begin{center}
\end{center}
\caption{
The $ \alpha $ dependence of (a) diagonal and (b) off-diagonal
elements of the generalized Fisher information matrix $G$
with $ \beta=0.5 $ and $I=1.0$.
The $\beta$ dependence of (c) diagonal and (d) off-diagonal
elements of $G$
with $\alpha=0.5$ and $I=1.0$.
The $I$ dependence of (e) diagonal and (f) off-diagonal
elements of $G$
with $\alpha=0.5$ and $\beta=0.5$.
In (a), (c) and (e), solid, dashed and chain curves denote
$g_{aa}/5$, $g_{bb}$ and $g_{cc}$, respectively:
in (b), (d) and (f), solid, dashed and chain curves denote
$g_{ab}$, $g_{bc}$ and $g_{ac}$, respectively.
}
\label{figH}
\end{figure}

\begin{figure}
\begin{center}
\end{center}
\caption{
The time-dependent distribution $p(x,t)$ in response to
an applied pulse input given by Eq. (\ref{eq:E1}):
solid and dashed curves express the results
calculated by the $q$-moment approach
and the PDE method, respectively 
($\lambda=1.0$, $\alpha=0.5$ and $\beta=0.5$).
Curves are consecutively shifted downward
by 0.25 for a clarity of the figure.
}
\label{figG}
\end{figure}

\begin{figure}
\begin{center}
\end{center}
\caption{
(a) The time dependence of $\mu_q$ (the bold solid curve) and 
$\sigma_q^2$ (the thin solid curve) 
calculated by Eqs. (\ref{eq:C4}) and (\ref{eq:C5})
for an applied pulse input $I$ (the chain curve):
dashed curves denote the results of the PDE method:
$\sigma_q(t)^2$ is multiplied 
by a factor of five.
(b) The time dependence of $S_q$ calculated
by Eqs. (\ref{eq:F7}) and (\ref{eq:F3}) (the solid curve) 
and by the PDE method (the dashed curve)
($\lambda=1.0$, $\alpha=0.5$ and $\beta=0.5$).
}
\label{figA}
\end{figure}

\begin{figure}
\begin{center}
\end{center}
\caption{
(a) The time dependence of diagonal elements
of the generalized Fisher information matrix $G$
in response to a pulse input given by Eq. (\ref{eq:E1}):
$g_{aa}$ (the solid curve), $g_{bb}$ (the dashed curve) and
$g_{cc}$ (the chain curve), $g_{aa}$ being divided 
by a factor of five.
(b) The time dependence of off-diagonal elements of $G$:
$g_{ab}$ (the solid curve), $g_{bc}$ (the dashed curve) and
$g_{ac}$ (the chain curve)
($\lambda=1.0$, $\alpha=0.5$ and $\beta=0.5$).  
}
\label{figB}
\end{figure}

\begin{figure}
\begin{center}
\end{center}
\caption{
The time dependence of the inverse of 
the generalized Fisher information matrix, $h_{ij}$ [$=(G^{-1})_{ij}$],
in response to a pulse input given by Eq. (\ref{eq:E1}):
$h_{aa}$ (the solid curve), $h_{bb}$ (the dashed curve)
and $h_{cc}$ (the chain curve),
$h_{bb}$ being divided by a factor of twenty
($\lambda=1.0$, $\alpha=0.5$ and $\beta=0.5$).  
}
\label{figC}
\end{figure}

\begin{figure}
\begin{center}
\end{center}
\caption{
(a) The time dependence of $\mu_q$ (the bold solid curve) and 
$\sigma_q^2$ (the thin solid curve) 
calculated by Eqs. (\ref{eq:C4}) and (\ref{eq:C5})
for an applied sinusoidal input $I$ (the chain curve):
dashed curves denote the results of the PDE method:
$\sigma_q(t)^2$ is multiplied 
by a factor of five.
(b) The time dependence of $S_q$ calculated
by Eqs. (\ref{eq:F7}) and (\ref{eq:F3}) (the solid curve) 
and by the PDE method (the dashed curve)
($\lambda=1.0$, $\alpha=0.5$ and $\beta=0.5$).
}
\label{figF}
\end{figure}

\begin{figure}
\begin{center}
\end{center}
\caption{
(a) The time dependence of the diagonal elements
of the generalized Fisher information matrix $G$
in response to a sinusoidal input given by Eq. (\ref{eq:E2}):
$g_{aa}$ (the solid curve), $g_{bb}$ (the dashed curve) and
$g_{cc}$ (the chain curve), $g_{aa}$ being divided 
by a factor of five.
(b) The time dependence of the off-diagonal elements of $G$:
$g_{ab}$ (the solid curve), $g_{bc}$ (the dashed curve) and
$g_{ac}$ (the chain curve)
($\lambda=1.0$, $\alpha=0.5$ and $\beta=0.5$).  
}
\label{figE}
\end{figure}

\begin{figure}
\begin{center}
\end{center}
\caption{
The time dependence of 
the generalized Fisher information matrices of (a) $g_{xx}$ (the chain curve) 
and $g_{\mu\mu}$ (the solid curve), and
(b) $g_{\gamma \gamma}$ (the solid curve) 
in response to a pulse input
given by Eq. (\ref{eq:E1})
($\lambda=1.0$, $\alpha=0.5$ and $\beta=0.5$) (see text). 
}
\label{figD}
\end{figure}
\end{document}